# Defining definition: a Text mining Approach to Define Innovative Technological Fields

**Giordano V.[a], Cervelli E.[a], Chiarello F.[a]**
giordano.vito94@gamil.com
elenacervelli2@gmail.com
filippochiarello.90@gmail.com

[a] *Department of Energy, Systems, Territory and Construction Engineering, University of Pisa*

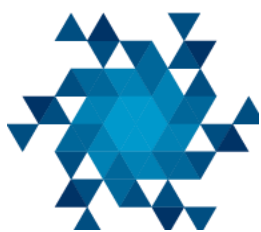

R&D MANAGEMENT CONFERENCE 2019 - DATA SCIENCE FOR INNOVATION

## Abstract

One of the first task of an innovative project is delineating the scope of the project itself or of the product/service to be developed. A wrong scope definition can determine (in the worst case) project failure. A good scope definition become even more relevant in technological intensive innovation projects, nowadays characterized by a highly dynamic multidisciplinary, turbulent and uncertain environment. In these cases, the boundaries of the project are not easily detectable and it's difficult to decide what it is in-scope and out-of-scope.

The present work proposes a tool for the scope delineation process, that automatically define an innovative technological field or a new technology. The tool is based on Text Mining algorithm that exploits Elsevier's Scopus abstracts in order to the extract relevant data to define a technological scope. The automatic definition tool is then applied on four case studies: Artificial Intelligence and Data Science.

The results show how the tool can provide many crucial information in the definition process of a technological field. In particular for the target technological field (or technology), it provides the definition and other elements related to the target.

# 1  Introduction

According to *Bryce* [3] the project scope is used to define the business problem and the opportunity to be. The scope should be clear and has to remain the same for the whole project. As *C. Cho* define [4], a poor scope definition is recognized by industry practitioners as one of the leading causes of project failure, adversely affecting projects in the areas of cost, schedule, and operational characteristics. For these reasons a well-defined scope is fundamental in project and it becomes very important in innovation process.

Innovative projects are characterized by an high degree of uncertainty. The risk linked to uncertainty (in terms of both probability and magnitude) become even more relevant in innovative projects in technological fields, characterized by a highly dynamic environment: multidisciplinarity, turbulence and uncertainty. [5] In these cases, the boundaries are not easily detectable and it's difficult to decide what it is in-scope and out-of-scope.

The present work demonstrates that it is possible to define new technological fields or technologies using text mining tools, in order to support the innovators and researchers in scope definition process within innovative projects. To construct a definition of a target innovative tech field, text mining techniques are applied to Elsevier's Scopus abstracts for extract relevant information that is useful in the scope definition process of target technological field:

1. definitions: A definition is a statement of the meaning of the term;
2. hyponyms: A hyponym of a term $x$ is a term $y$ included in a semantic field of term $x$.
3. hypernyms: A hypernym of a term $x$ is a term $y$ that includes in a its semantic field the term $x$.

The present work is structured as follow. In section **2** the relevant literature to understand our work is reported, in particular an explanation of what a definition is. In section **3** the developed methodology to develop a tool that aims to automatically define a tech field is presented. In section 4 the automatic definition tool is applied on two case studies: Artificial Intelligence and Data Science. Finally, in section **5** we discuss about the conclusion and the next steps of our work.

# 2  Literature in Brief

## 2.1 Defining Definitions

The purpose of a definition is to map the meaning of a term in order to provide the user with an understanding on what the term is about. *John, L. (1977)* [6] classifies the definitions in two large categories: intensional definitions and extensional definitions. *Roy T. Cook (2009)* explains the differences between these two types of definition [7] as:

- "An intensional definition gives the meaning of a term by specifying all the properties required to come to that definition, that is, the necessary and sufficient conditions for belonging to the set being defined".
- "An extensional definition defines by listing everything that falls under that definition".

In new technological fields, both categories of definitions are required to understand completely the field, but this work is focused on intensional definition, because this definition form can help to develop a set of rules for mining definitions from texts. Also, a list of elements under the meaning of technological field is provided, for attempt to give an extensional definition of tech field. A type of intensional definition is the *Genus differentia definition,* defined by *Parry, W. T., & Hacker*. In [8] is reported "Aristotle recognized only one method of real definition, namely, the method of genus and differentia, applied to defining real things, not words." In this section a *Genus differentia model* is discussed, based on *Parry, W. T., & Hacker, E. A. (1991)* book [8]. Our work is focused on *Genus differentia definition*.

Based on genus differentia model, the definition, from a more technical point of view, is an equivalence relation between two element definiendum and definiens:

$$x = y + z$$

$$x \leftarrow definiendum$$

$$= \leftarrow definitor$$

$$(y + z) \leftarrow definiens$$

**Definiendum** is the term, that we will be defined. **Definitor** is a part of definition that relate in an equivalence relation definiendum and definiens. **Definiens** is an expression that defines definiendum. A definiens is composed by a principal concept and by the enunciation of the features that distinguish the definiendum from the principal concept. The concept is a hypernym of the definiendum and it identifies the class of concept which includes the definiendum. Hypernyms are called genera. Distinctive features must allow to differentiate definiendum from genus and other cohyponym. Thus,

$$x = y + z$$

$$y \leftarrow genus$$

$$z \leftarrow set\ of\ distinctive\ features$$

$z$ is a summation of more component, each of one represent a single distinctive feature. This type of definition is called *Genus differentia definition* because it starts from part genus ($y$) and adds the features to $y$ to differentiates $x$ from other terms. *Genus differentia model* is useful to define the relation and properties among terms in a mathematical way. It is important to provide the users with the standard delineation of these terms and not to confuse between semantic world. The delineation of some semantic relation between the terms is fundamental to understand the work presented in this paper, in particular the definition of semantic field, hypernymy, hyponymy, synonymy is constructed starting from *Genus Differentia* definition. *Green R*. (2013) tries to define a subset of these terms in a mathematical way, but in a different manner from the one presented [9]. For the scope of this work, we prefer the forms described below, because they are based on *Genus differentia definition*; we will stick to one school of thought in order to provide the same point of view throughout the whole paper.

**Definition 1** - A **semantic field** is a set of words grouped semantically. For example, the semantic field of word organ is a set of word *{heart; liver; small; ...}*. Field semantic of term $y$, called $C_y$ is:

$$C_y = \{x \mid x = y + z\}$$

**Definition 2** - **Hypernymy** is a relationship that relates two terms $x$ and $y$, in which a term $y$, called hypernym, includes in its own semantic field other terms $x$, called hyponyms, that have a semantic field smaller than $y$. Thus:

$y$ *is* **hypernym** of $x$ $\iff x = y + z \land C_y \ni x$

Consequently $C_y \supset C_x \land \#C_x < \#C_y$.

In other words, a hypernym is a term that indicate a lexical unit of meaning more generic and extensive than one or more lexical units, that are included in the semantic field of the hypernym. For example, reptile is hypernym of lizard. The contrary of hypernym is hyponym.

**Definition 3** - **Hyponymy** is a relationship that relates two terms *x* and *y*, in which a term *x*, called hyponym, is included in semantic field of other term *y*, called hypernym, that has a semantic field which is more extensive than *x*. Thus:

$$x \text{ is } \textbf{hyponym} \text{ of } y \iff x \in C_y$$

Consequently $C_x \subset C_y \land \#C_x < \#C_y$. For example, sunflower is a hyponym of flower.

**Definition 4** - **Synonymy** is a relationship that relate two terms *x* and *y*, where $x \neq y$, but they have the same definiens. Thus:

$$x \text{ is } \textbf{Synonymous} \text{ of } y \text{ and vice versa} \iff \{x = h + z \; y = h + z$$

Consequently $x \in C_h \land y \in C_h$. For example, flask is a Synonymous of balloon.

## 2.2   Text mining techniques

Text Mining a field of research which helps in getting relevant information from unstructured textual data. It is an interdisciplinary field which draws on information retrieval, data mining, machine learning, statistics and computational linguistics. Since most information, over 80%, is stored as text, text mining is believed to have a high commercial potential value [10].

The problem introduced by text mining is obvious: natural language was developed for humans to communicate with one another and to record information, and computers are a long way from comprehending natural language. Most advanced text mining software use sophisticated Natural Language Processing (NLP) algorithms. Natural language processing (or NLP) is a component of text mining that performs a special kind of linguistic analysis that essentially helps a machine "read" text [11].

In the present work, the most important NLP tools is universal POS tagging: it marks the core part-of-speech categories and to distinguish additional lexical and grammatical properties of words, use the universal features. The used systems was developed in CoNLL, that stands for Conference on Natural Language Learning and is the SIGNLL's (Special Interest Group on Natural Language Learning) yearly meeting [12]. In particular we used the R package udpipe [13], that uses a revised version of the CoNLL-X format called CoNLL-U [14]. The most relevant aspect of CoNLL-U annotation, Text chunking is borrowed from Natural Language Processing. The activity allows to divide sentences into nonoverlapping segments [15] and is done in unsupervised mode by directly dividing sentences into phrases using linguistics or statistics [16].

The second relevant techniques used in this article is Named-entity recognition (NER) , that is a subtask of information extraction that seeks to locate and classify named entity mentions in unstructured text into pre-defined categories such as the person names, organizations, locations [17].

These Text Mining techniques has been used in this work to manipulate the Scopus abstracts in order to extract the relevant data from it, such as definitions, hypernyms and hyponyms. We used only the abstracts rather than the entire body of each article because having access to all the articles on a given topic is often difficult, while the abstracts can be sources for massive information analysis with very low cost.

## 3 Methodology

The methodology to build a tool that aims to automatically define a technological field has been divided in three phases:

1. **Rules construction**: in this phase a set of rules has been designed to extract from abstracts of scientific papers the definitions of a technological field (or other relevant elements that can be useful in definition process). The rules have been implemented in the form of regular expression. A regular expression is a sequence of strings, used to identify text data that follow the regularity sought [18]. All analytic process has been developed on software Rstudio.
2. **Rules Validation**: the capability of the constructed rules to identify relevant element in the text has been evaluated through a set of observations (tech field or technology).
3. **Definitions analysis**: In this final phase the methodology has been applied to case studies. In particular, the definitions, hypernyms and hyponyms of the case studies have been mined from the Scopus database of scientific articles. For each case studies we: (i) analysed the frequency distribution of genera used in extracted definitions; (ii) analysed the frequency distribution of distinctive used in extracted definitions; (iii) analysed the co-occurrence frequency of distinctive used in extracted definitions; (iv) constructed an ontology of the tech field using extracted hyponyms and hypernyms. An ontology is a classification of existing concepts in the technological field. Finally, the results of different case studies have been compared among them.

The proposed process is shown in *figure 1*. The following section describes in depth each task of presented methodology.

### 3.1 Construct top-down rules

To construct the rules for extract definitions, hypernyms and hyponyms, a hybrid theoretical-empirical approach has been developed, called respectively top-down rules and bottom-up rules. The top-down rules have been constructed starting from sources in the literature concerning the formulation of a formal definition. The theory explained in section **2.1** about the *Genus differentia definition* has been used in this task as a pillar to generate all rules. Starting from the Wikipedia page of *definition* and from other sources mentioned in section **2.1** (such as [6], [7]) and from [20], a complete list is provided joined with the exploration of related documents. Thus, the list of sources establishes the theoretical base for the definition process and to perform the construct top-down rules task.

### 3.2 List random technologies

The bottom-up rules have been formulated by analyzing more than 600 definitions of technologies extracted from Wikipedia. Starting from the Wikipedia page of *List of emerging technologies* [19], we extract the hyperlinks to list random technologies. The repetition of this process with the extracted technologies enlarges the list of words. Though, since consistency of analysis is crucial in bottom-up processes, the starting technologies list must be large enough; we aim to an extraction of more than 600 definitions. If the number were to fall below 600, we would require the enlargement of the technology list.

To mine definitions on Wikipedia, a regular expression is built using the wished technology term. This task is based upon the hypothesis that the Wikipedia page of a term starts with the definition of the same term.

### 3.3 Analyze definitions

In this step, extracted definitions of listed technologies are screened. The dataset of definitions of each term belonging to list of technologies has been analysed to build a set of bottom-up rules. In particular, the observed element in each definition have been: (i) definiendum; (ii) definitor; (iii) genus, (iv) words between definiendum and definitor. Finally, this task aims to obtain a list of empirical rules in order to extract definitions, hyponyms, hypernyms and synonyms from text, in specific instance we will try to mine these relevant data from Scopus abstracts.

The rules have been then classified into families, based on information extracted from the texts, and in classes, grouping the rules of each family according to multiple criteria. *Table 1* shows a part of the 36 rules identified in Rules construction phase. All established rules are available in [21]. The families of the rules are:

- Definition: rules with the function of extracting case study definitions;
- Hyponym: rules with the function of extracting hyponyms and hypernyms of the observations.

The possible classes, according to which a rule can be classified, are:
- Definition starting: rules that indicate the sequence of words with which a definition can begin;
- Definitor: rules that indicate the definitors that can be used to extract the definitions from a text;
- Definitor following: rules that indicate the characters that follow the definitor within a definition;
- Genus structure: rules that indicate the Part of Speech constituting the genus of a definition;
- Complete definition: rules that allow you to extract a complete definition from a text;
- Hyponym core: rules referring to the main elements that allow identifying a hyponym in the abstracts of Elsevier's Scopus;
- Hyponym structure: rules that indicate the Part of Speech constituting the hyponyms;
- Synonymous structure: rules that refer to the structure of synonyms of the term to be defined within the definitions.

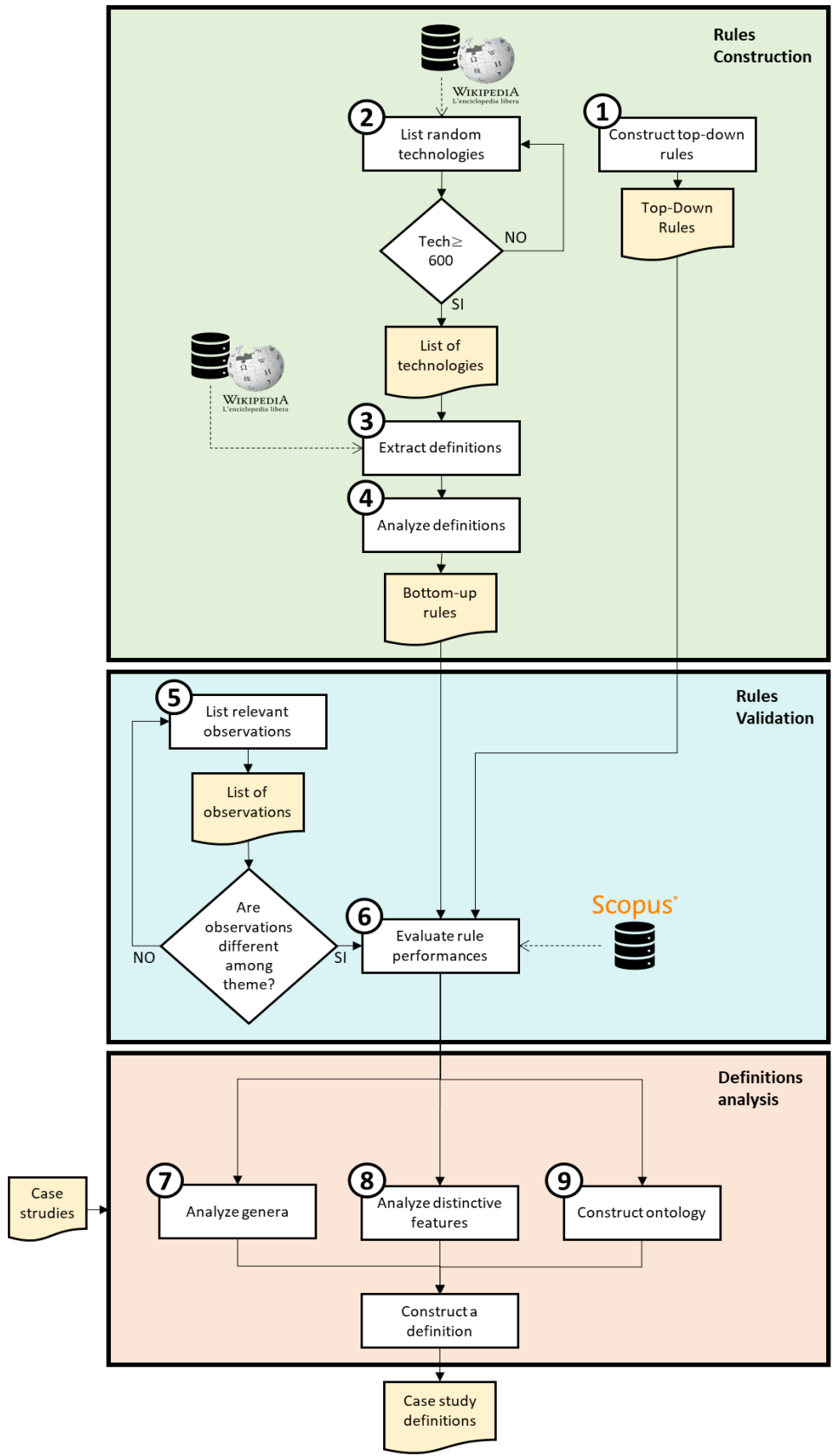

*Figure 1 - Process flow diagram of the proposed methodology for build a tool that aims to automatically define an innovative tech field. The diagram contains the representation of the documents and the operations performed on them.*

## 3.4 List relevant observations

In order to test and evaluate the constructed rules, the relevant observations must be listed. An observation is a technology or technological field on which the rules can be applied. In order to validate the rules choosing distant observations is crucial. For the scope of this work, distance is defined on the basis of three main layers:

   a. Interest from scientific community in technology or technological field;
   b. Time when term relative to technology or technological field was coined;
   c. Space spread of the term. It means for technologies the spread in subject areas in which the technology is applied and for technological fields, the spread in subject areas in which the technological field is studied.

To measure how distant is one technology from the others, we can retrieve indicators of these three variables. The results are, respectively:

   a. Number of technology (or technological field) papers on Elsevier's Scopus;
   b. The year in which the first paper relative to said technology was published on Elsevier's Scopus;
   c. Gini index of the subject area distribution in technology.

The selection of technology and technological field starts with idea to provide a vision of the emerging technological fields in these years. In order to ensure consistency of the rules validation process, the number of observations must be at least 10. The observations used to rules validation process are: *Data Science*, *Artificial Intelligence*, *Industry 4.0*, *Financial Technology*, *Machine Learning*, *Automated Guided Vehicle*, *Business Intelligence*, *Precision Agriculture*, *Knowledge Management, Chatbot, Natural Language Processing*, *Unmanned Aerial Vehicle*, *Internet of Things* e *Deep Learning*.

| Rule | Type | Family | Class | Example |
|---|---|---|---|---|
| **Definiendum** | Top-down | Definition | Definition starting | **Artificial intelligence** is a branch of computer science. |
| **Article + definiendum** | Top-down | Definition | Definition starting | **The Artificial intelligence** is a branch of computer science. |
| **be** | Bottom-up | Definition | Definitor | Artificial intelligence **is** a branch of computer science. |
| **be + defined as** | Bottom-up | Definition | Definitor | Artificial intelligence **is defined as** a branch of computer science. |
| **refer to** | Bottom-up | Definition | Definitor | Artificial intelligence **refers to** a branch of computer science. |
| **Definiendum** + *zero or one word* + **such as** + **hyponym** | Top-down | Hyponym | Hyponym core | Recently, **artificial intelligence techniques such as fuzzy logic**, neural networks and genetic algorithms are used to solve various problems in trading. |

*Table 1 – A part of 36 constructed rules. For each rules the type (top-down or bottom-up), the family, the class has been specified and an example is provided to ensure the o ensure understanding of every rule.*

## 3.5 Evaluate rule performances

For each rule we apply the prototype with the aim of extracting definitions and hyponymies from all the observed Scopus' abstracts. The results have been manually analysed to understand if rules could be useful for the tool construction. Rules belonging to the same class have been evaluated and ranked to decide which of them could represent the basis for the tool (output of the present work). The ranking took into consideration the following aspects:

- Number of relevant results identified by the prototype
- Precision, given by the number of relevant results on the total number of phrases extracted by the prototype

When rules build a definition or identify hyponymies, the results are relevant. The rules having at least a precision of about 65% and a high number of relevant results could represent the basis for the tool's behaviour. If some rules are exclusive, so if they can't both belong to the Automatic definition tool, only the rule among them having the higher number of relevant results has been chosen. An example of rules selection is given by *table 2*: two rules belonging to Definitor class have been compared. Since they were not exclusive, the possibilities analysed were:

- Use only definitor {"is"};
- Use only definitor {"refers to"};
- Use both of them {"refers to","is"}.

| Observation | {"Is"} | | | {"Refers to"} | | | Both | | |
|---|---|---|---|---|---|---|---|---|---|
| | Number of retrieved results | Number of relevant results | Precision | Number of retrieved results | Number of relevant results | Precision | Number of retrieved results | Number of relevant results | Precision |
| Data Science | 32 | 25 | 0,781 | 1 | 1 | 1,00 | 33 | 26 | 0,788 |
| Artificial intelligence | 168 | 102 | 0,607 | 4 | 3 | 0,75 | 172 | 105 | 0,610 |
| Industry 4.0 | 82 | 69 | 0,841 | 6 | 6 | 1,00 | 88 | 75 | 0,852 |
| Financial Technology | 3 | 3 | 1,000 | 1 | 1 | 1,00 | 4 | 4 | 1,000 |
| Machine learning | 233 | 160 | 0,687 | 5 | 5 | 1,00 | 238 | 165 | 0,693 |
| Automated Guided Vehicle | 2 | 2 | 1,000 | 0 | 0 | 0,00 | 2 | 2 | 1,000 |
| Business intelligence | 44 | 37 | 0,841 | 1 | 1 | 1,00 | 45 | 38 | 0,844 |
| Precision agriculture | 55 | 44 | 0,800 | 4 | 4 | 1,00 | 59 | 48 | 0,814 |
| Knowledge management | 222 | 110 | 0,495 | 11 | 10 | 0,91 | 233 | 120 | 0,515 |
| Chatbot | 18 | 18 | 1,000 | 0 | 0 | 0,00 | 18 | 18 | 1,000 |
| Natural language processing | 38 | 28 | 0,737 | 1 | 1 | 1,00 | 39 | 29 | 0,744 |
| Unmanned aerial vehicle | 34 | 26 | 0,765 | 2 | 2 | 1,00 | 36 | 28 | 0,778 |
| Internet of things | 421 | 281 | 0,667 | 71 | 70 | 0,99 | 492 | 351 | 0,713 |
| Deep learning | 199 | 106 | 0,533 | 4 | 4 | 1,00 | 203 | 110 | 0,542 |
| **MEAN** | | 72 | 0,652 | | 8 | 0,97 | | 80 | 0,673 |

*Table 2 – Rules validation example: the comparison between {"Is"} and {"Refers to"}, two rules belong to Definitor Class.*

Column "number of relevant results" contains the mean number of results obtained thanks to the rule's application on total observations. Columns "precision" indicates the mean of all observations' own precision, weighted on extracted sentences, with the aim of reducing importance of observations having high precision but few extracted sentences at the same time. In *table 2* is shown how the joint use of both rules involves better performances (higher number of extracted sentences and higher tool's precision). For this reason, both rules have been chosen to represent the basis of the Automatic definition tool. The validation process has been reiterated for all rules' classes. The global precision of the tool reached 68% and it can extract for each technology field about 80 definitions and 240 hyponymies and relative hypernymies.

### 3.6 Analyse genera

In the next sections the analysis of case studies is faced, as we shown in *figure 1*. The validate rule have been implement in R in two functions that aim to scrap the definitions and the hyponyms with related hypernyms from the Elsevier's Scopus abstracts for each case studies: Artificial Intelligence, Industry 4.0, Data Science and Internet of things. For each technological field the definitions, hyponyms and hypernyms have been mining in order to build a good quality definition of the case study. From each extracted definition for a specific tech field, the genus has been mining with the constructed rules described in section **3.5** and the frequency distribution of genera will be analysed. To perform this process, a text mining tools are used to capture the genera, in fact the constructed rules are based on theory of chunking, that ensure to mine all words that compose a genus in a definition and not a part of this. The purpose of this task is identified the genera to insert in our definitions of the observed tech fields.

### 3.7 Analyse distinctive features

For each case study, the analysis of distinctive features is performed observing: (i) the frequency distribution of distinctive features used in definitions of observed tech fields; (ii) the frequency co-occurrence of distinctive features used in definitions of observed tech fields. The distinctive features have been extracted from each definition with the constructed rules in section **3.5**. A distinctive feature is a word or a chunking of words, that are relevant in the definition, so the stopwords have been removed to clean the definitions from noise.

In the distribution of distinctive features, the frequency of each distinctive feature has been calculated as the number of times that the specific distinctive feature is used in a definition over the definition total number of the observed technological field. For example, the distinctive feature *knowledge* appears in 30 % of the Data Science technological field definitions.

Meanwhile, the co-occurrence refers to a frequency of occurrence of two distinctive features from a same definition, for instance the terms *knowledge* and *extract* in Data Science definitions appear together in the 10 % of the cases. The analysis performed in this section are useful to understand the main distinctive features to insert in our definitions of the observed tech fields.

### 3.8 Constructed ontology

The scope of this process is to create an ontology of the observed tech field using the hypernyms and hyponyms extracted from Scopus abstracts. Also in this case, the extraction of these elements is possible thanks to the constructed rules in section **3.5**.

Finally, for Artificial Intelligence and Data Science, a definition will be building using the genera, distinctive features analysis and the ontology. The two case studies will be compared among them.

## 4 Results

The following section describes the performance of automatic definition process on two different case studies: Artificial Intelligence and Data Science. Each tech field is analysed in each section of this part. Finally, in section **4.3** the comparison of all technological fields will be shown.

### 4.1 Case studies results: Artificial Intelligence

The extracted definitions of Artificial Intelligence are 107. An example of these is shown in *table 3*. Furthermore, the hypernyms and hyponyms extracted with the tool are 605 and a part of these is shown in *table 4*.

| Scopus ID | Definition | Genus | Distinctive features |
|---|---|---|---|
| 2-s2.0-85054690028 | artificial intelligence is a branch of computer science connects classifies differentiates and elaborates the domains of learning in neural network a paradigm shift is using in the construction of knowledge | branch of computer science | connect, classifies differentiates, elaborate, domain, learning, neural network, paradigm shift, construction, knowledge |
| 2-s2.0-85046415420 | artificial intelligence is the ability of a computer to perform the functions and reasoning typical of the human mind | ability of a computer | perform, function, reason, typical, human mind |
| 2-s2.0-85055517085 | artificial intelligence is a science and technology to study of the law of human intelligence activities | science | technology, study, law, human intelligence activities |
| 2-s2.0-85051252856 | artificial intelligence ai is a branch of computer science that deals with the problemsolving by the aid of symbolic programming | branch of computer science | deal, problemsolving, aid, symbolic programming |
| 2-s2.0-85045918452 | artificial intelligence is the study of intelligent machines or intelligent agents or robots | study of intelligent machines | intelligent agents, robot |

*Table 3 - An example of extracted definitions of **Artificial Intelligence** from Scopus abstracts. In column "Scopus ID" the identification code of the article is shown, referred to the abstract in which definition is contained. In column "Definition", the extracted definition is reported. In column "Genus", the genus used in definition is shown. In column "Distinctive Features", the distinctive features used in definition are reported, each distinctive feature has been separated among others with comma.*

| Scopus ID | Sentence | Hyponym | Hypernym |
|---|---|---|---|
| 2-s2.0-85048852027 | deep neural networks (dnns) have been recently achieving state-of-the-art performance for many artificial intelligence (ai) applications such as computer vision, image recognition, and machine translator. | image recognition | application |
| 2-s2.0-85048852027 | deep neural networks (dnns) have been recently achieving state-of-the-art performance for many artificial intelligence (ai) applications such as computer vision, image recognition, and machine translator. | machine translator | application |
| 2-s2.0-85055695501 | ai driven applications such as autonomous vehicles, medical diagnostics, conversational agents etc. are becoming a reality. | medical diagnostics | application |
| 2-s2.0-84934923322 | the literature survey is organized based on different artificial intelligence techniques such as fuzzy logic, genetic algorithms, neural networks, game theory, reinforcement learning, support vector machine, case-based reasoning, entropy, bayesian, markov model, multi-agent systems, and artificial bee colony algorithm. | bayesian | technique |
| 2-s2.0-84929072533 | further advances in ai technologies such as natural language comprehension and image recognition will only increase surveillance powers. | natural language comprehension | technology |
| 2-s2.0-85046720779 | we conduct a holistic, systematic literature review using artificial intelligence technologies such as information retrieval, text mining and supervised learning, side-by-side with manually reading of many relevant articles. | supervised learning | technology |

*Table 4 - An example of extracted hyponyms and hypernyms of **Artificial Intelligence** from Scopus abstracts. In column "Scopus ID" the identification code of the article is shown, referred to the abstract in which hypernyms and hyponyms are contained. In column "Sentence", the sentence is reported, in which hyponyms and hypernyms are contained. In column "Hyponym", the hyponym extracted from sentence is shown. In column "Hypernym", the hypernym of hyponym extracted from sentence is reported.*

In *figure 2* the analysis of the Artificial Intelligence definitions-set is shown, in particular:

a. The analysis of the frequency distribution of genera used in definitions-set;
b. The analysis of the frequency distribution of distinctive features used in definitions-set;
c. The analysis of the frequency of co-occurrences of distinctive features: the frequency of co-occurrences represents the number of times a pair of distinctive features appears together in the set of definitions.
d. The ontology: a classification of concept in Artificial Intelligence has been constructed thanks to the extracted hypernyms and hyponyms of Artificial

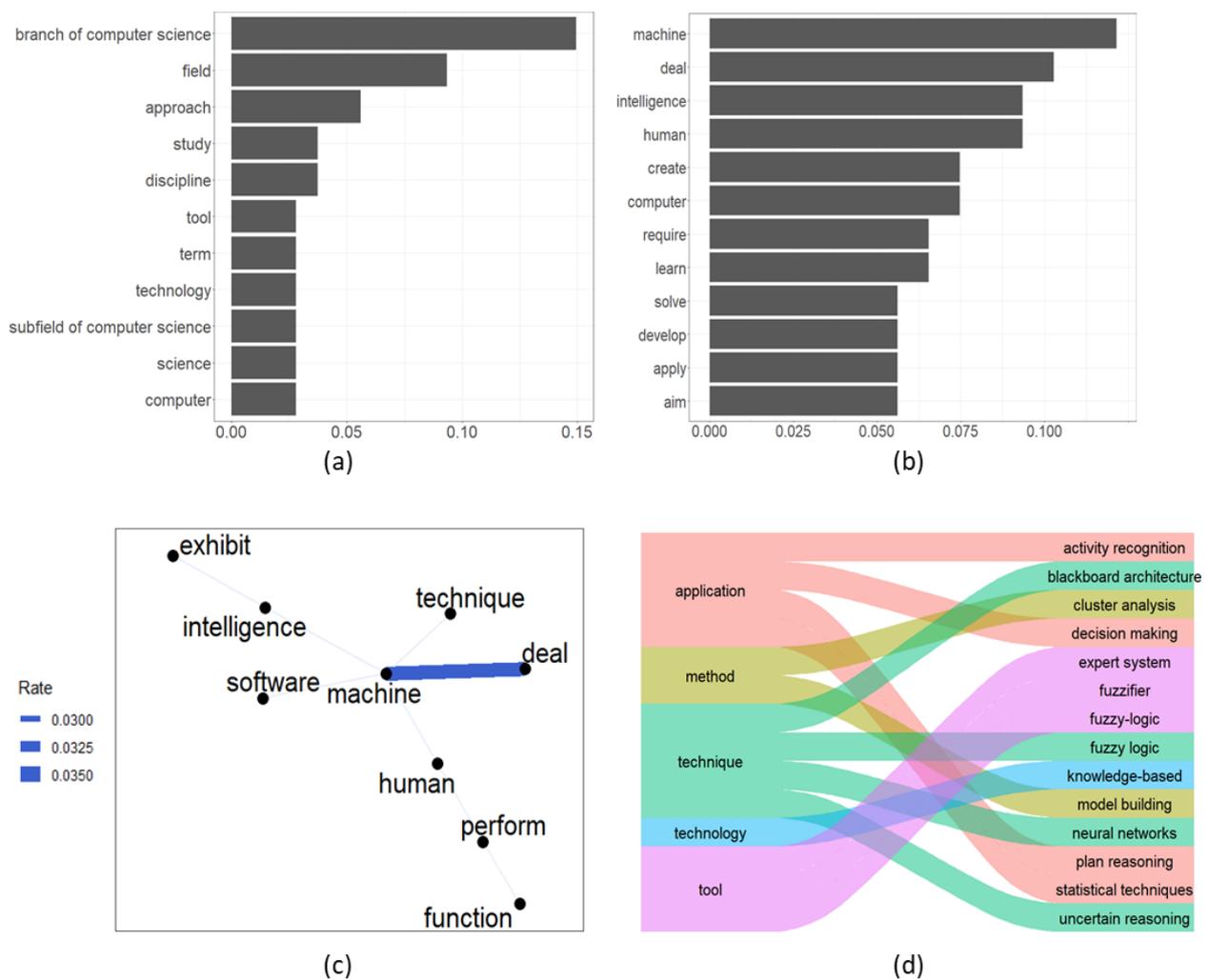

Intelligence.

*Figure 2 – Analysis of **Artificial Intelligence** definitions-set. **(a)** The frequency distribution of genera used in Artificial Intelligence definitions-set. **(b)** The frequency distribution of distinctive features used in Artificial Intelligence definitions-set. **(c)** The frequency of co-occurrence of distinctive features used in Artificial Intelligence definitions-set. **(d)** The ontology of Artificial Intelligence.*

To verify the robustness of the analysis some aspects have been analysed and plotted. *Figure 2a* communicates which genera are more used in Artificial Intelligence definitions, e.g. the chunk "branch of computer science" has an occurrence probability of 15%. The maximum occurrence probability is low: this means there isn't a convergence towards a dominant definition. The distribution of distinctive features and the co-occurrence of distinctive features in definitions are represented in *figure 2b* and *figure 2c*: confusion in terms is also evident in this area. The distinctive feature "machine" appears in about 12% of definitions (low percentage). The terms "machine" and "deal" in Artificial Intelligence definitions appear together in about 3,3% of cases (weak bond). No other terms seem to be strongly connected, resulting in confusion and uncertainty in defining the technological field. In *figure 2d* an ontology of Artificial Intelligence is shown: some hypernyms and hyponyms and relations among them are shown. Coverage is partial for synthesis reason but it's possible to build a richer ontology.

Using the outputs produced by the Automatic definition tool, the definition of Artificial Intelligence has been provided. We have taken into consideration *figure 2a* to identify the most suited genera, *figure 2b* and *figure 2c* to select and combine the better distinctive features and, at the end, *figure 2d* to generate an example.

> **Def**. *Artificial intelligence is the branch of computer science that develops machines with intelligence like human.*
>
> **E.g.** *Some possible applications of Artificial intelligence are in activity recognition, decision making, plan reasoning and statistical techniques tasks.*

## 4.2 Case studies results: Data Science

The extracted definitions of Data Science are 26. An example of these has shown in *table 5*. Otherwise, the hypernyms and hyponym extracted with the tool are 27 and a part of these is shown in *table 6*.

| Scopus ID | Definition | Genus | Distinctive features |
|---|---|---|---|
| 2-s2.0-85059974598 | data science is an interdisciplinary field that uses scientific methods from computer science and statistics to extract insights or knowledge from data in a specific domain | field | scientific methods, computer science, statistic, extract insights, knowledge, data, specific domain |
| 2-s2.0-85060791688 | data science is an interdisciplinary field that is very much like data mining and knowledge discovery in databases kdd involving the analysis of data to make useful inferences and deduction | field | data mining, knowledge discovery, database, kdd, involve, analysis, data, deduction |
| 2-s2.0-85062206341 | data science is a successful study that incorporates varying techniques and theories from distinct fields including mathematics computer science economics business and domain knowledge | study | incorporate, varying, technique, theory, distinct fields, include, mathematics computer science economics business, domain knowledge |
| 2-s2.0-85050493770 | data science is an interdisciplinary field that extracts insights from data through a multistage process of data collection analysis and use | field | extract, insights, data, multistage process, data collection analysis |
| 2-s2.0-85046758297 | data science is a hybrid of multiple disciplines and skill sets draws on diverse fields including computer science statistics and mathematics encompasses topics in ethics and privacy and depends on specifics of the domains to which it is applied | hybrid of multiple disciplines | skill, set, draw, diverse fields, include, computer science statistics, mathematics, encompass, topic, ethic, privacy, depend, specifics, domain, apply |

*Table 5 - An example of extracted definitions of **Data Science** from Scopus abstracts. In column "Scopus ID" the identification code of the article is shown, referred to the abstract in which definition is contained. In column "Definition", the extracted definition is reported. In column "Genus", the genus used in definition is shown. In column "Distinctive Features", the distinctive features used in definition are reported, each distinctive feature has been separated among others with comma.*

| Scopus ID | Sentence | Hyponym | Hypernym |
|---|---|---|---|
| 2-s2.0-85045124637 | © 2018 elsevier ltd feature selection is a crucial procedure in data science tasks such as classification, since it identifies the relevant variables, making thus the classification procedures more interpretable, cheaper in terms of measurement and more effective by reducing noise and data overfit. | data overfit | task |
| 2-s2.0-85046953082 | we demonstrate this approach in iris, an agent that can perform open-ended data science tasks such as lexical analysis and predictive modeling. | lexical analysis | task |
| 2-s2.0-85046953082 | we demonstrate this approach in iris, an agent that can perform open-ended data science tasks such as lexical analysis and predictive modeling. | predictive modeling | task |
| 2-s2.0-85049357068 | data science methodologies, such as autoencoding and text mining, were adapted to identify candidate gene sets that distinguish different types of cells in the central nervous system. | autoencoding | methodology |

*Table 6 - An example of extracted hyponyms and hypernyms of **Data Science** from Scopus abstracts. In column "Scopus ID" the identification code of the article is shown, referred to the abstract in which hypernyms and hyponyms are contained. In column "Sentence", the sentence is reported, in which hyponyms and hypernyms are contained. In column "Hyponym", the hyponym extracted from sentence is shown. In column "Hypernym", the hypernym of hyponym extracted from sentence is reported.*

In *figure 3* the analysis of the Data Science definitions-set is shown.

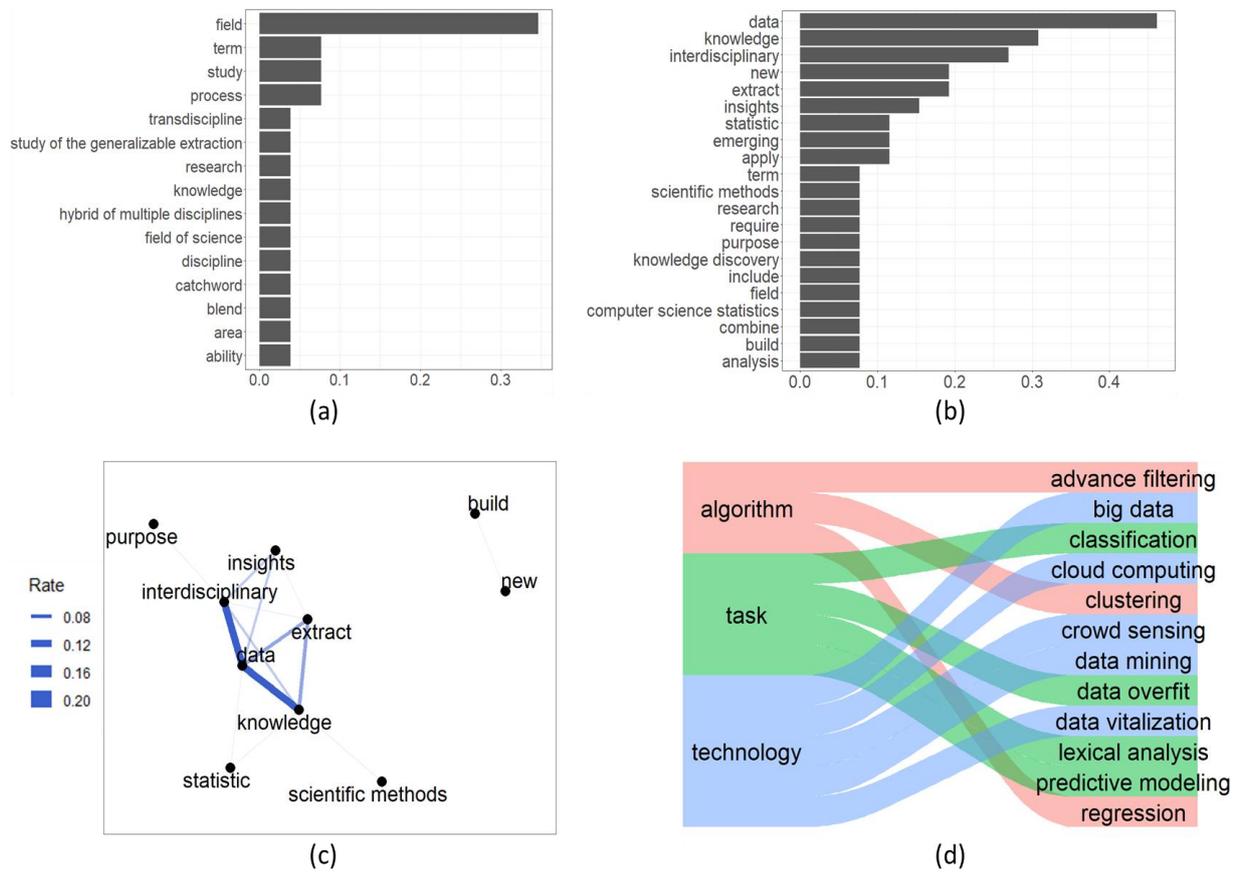

*Figure 3 – Analysis of **Data Science** definitions-set. (**a**) The frequency distribution of genera used in Data Science definitions-set. (**b**) The frequency distribution of distinctive features used in Data Science definitions-set. (**c**) The frequency of co-occurrence of distinctive features used in Data Science definitions-set. (**d**) The ontology of Data Science.*

*Figure 3a*, *figure 3b* and *figure 3c* show a stronger convergence in this Data Science field then AI. In fact, distribution of genera has a long tail and "field" term is used in more than one on three definition, occurrences of the most popular distinctive features are high, and terms are strongly connected.

Data science definition and example are extracted in the same way of the previous case study.

**Def.** *Data science is an interdisciplinary field with purpose to extract knowledge from data.*

**E.g.** *Some technologies used in Data science are: Data mining, Big data, Cloud computing.*

### 4.3 Case studies comparison

In this section a comparison between two analyzed case studies has been presented, in terms of convergence in the delineation process of tech field from scientific community. To compare the cases the network analysis has been performed. Each definition of Artificial Intelligence and Data Science has been represented as a vector based on the words found in the definition. To construct the network analysis the definitions have been represented in a graph, where a node is a definition of Artificial Intelligence or Data Science and an arc between two definitions represents the link between both definitions. The strength of link dependens on the number of words in common to both definitions.

The hypothesis is: if a technological field is old, then its definitions in literature tend to converge to the same meaning. On the other hand, a new technological field will be fuzzily defined, because it has not reached a level of maturity in order to have a commonly agreed definition. The hypothesis has been validated with some case studies in [21].

In our case the analysis of Artificial intelligence compared to Data science has been performed. Artificial intelligence first emerged in the 80' and then evolved at the beginning of the 21° century with neural networks. This has resulted in a lesser cohesion of the definitions as can be seen in the network diagram (*Figure 4*) where there are various clusters. On the other, data science tends to be more cohesive in terms of definitions even though it has a new paradigm.

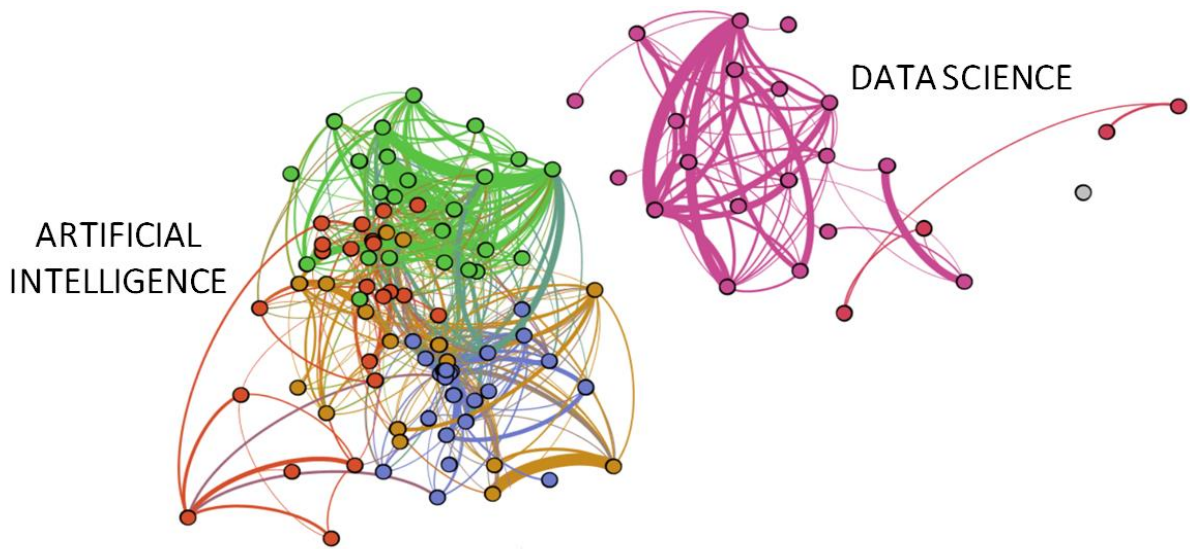

*Figure 4 – Cluster analysis of **Data Science** definitions-se and **Artificial Intelligence**. A node represents a definition extracted from Scopus related to the technological fields. The thickness of an arc between two definitions is proportional to the words found in both definitions.*

## 5 Conclusion

In the present paper we demonstrated that it is possible to main definition of innovative technologies or technological fields from the abstracts of scientific papers. We developed a methodology to solve this problem, and successfully applied the methodology to two case studies (Artificial Intelligence and Data Science). The results have been compared between the two case studies, showing also how our tool is able to identify fuzzy-defined technological fields.

A first next step is to understand if it is possible to extract entities that are similar to technologies such as methods or algorithms, using the presented methodology. The tool can be slightly modified in order to extract these other entities. Furthermore, other possible rules could be implemented in *Automatic definition tool* to enhance its precision and recall. Other sources can be mined for definition (e.g. patents or twitter) to enhance the ability of the tool to have a broad vision of a technological field. Finally, the tool has been designed to help the innovators and researchers in scope definition process, thus we want to evaluate in an field experiment if it is true, to assess the efficacy of the developed tool.